\documentclass[a4paper,twoside]{article}
\newcommand{\affil}[1]{$^{\rm #1}$}
\input{./psfig}
\date{} %Please leave the date blank
\title{\large\bf\flushleft Modelling early-type stars in eclipsing binaries of open
clusters: a new method for age determination from ratio of radii}
\author{\parbox{\textwidth}{\flushleft
\vspace{-0.5cm}
{\small \affil{}\,Ege University, Department of Astronomy and Space Sciences, Bornova, 35100 \.Izmir, Turkey}\\
{\small \affil{}\,Email: mutlu.yildiz@ege.edu.tr}}}
\begin{document}
\twocolumn[
\begin{changemargin}{.8cm}{.5cm}
\begin{minipage}{.9\textwidth}
\vspace{-1cm}
\maketitle
%
%
%%%%%%%%%%%%%     ABSTRACT    %%%%%%%%%%%%%
%Abstract of no more than 200 words here.
\small{\bf Abstract:}

Binary systems, in particular eclipsing binaries, are essential sources 
of our knowledge of the fundamental properties of stars.  The ages of binaries, 
members of open clusters, are constrained by their own fundamental properties 
and by those of the hosting cluster. The ages of eleven open clusters are here found by constructing 
models for the components of twelve eclipsing binaries. The difference between 
the ages we found and the ages of the clusters derived from isochrone fitting is up to 
40 \%. For the binary system V497 Cep in NGC 7160, the difference is about 100  \%. 
Binary systems
whose primary component is about 
to complete main-sequence life time, such as V453 Cyg and V906 Sco, are the most suitable systems for age 
determination.  Using model results for these stars, we derive an expression 
for sensitive and uncomplicated relative age determination of binary systems (age divided by 
the main-sequence life time of the primary star).
The expression is given as logarithm of radii ratio divided by a logarithm of mass ratio. 
Two advantages of this 
expression are that (i) it is nearly independent of assumed chemical 
composition of the models because of the appearance of the ratio of radii, and 
(ii) the ratios of radii and masses are observationally much more precise than their { absolute} values.
We also derive another expression using luminosities rather than radii and compare results.


\medskip
\medskip
\end{minipage}
\end{changemargin}
]
\small

\section{Introduction}
Nature is full of signals displaying its past and present structure.
Astrophysics is extremely specialized in detecting and extracting such signals from
very noisy backgrounds. In this context, research on deep regions of stars is a real adventure.
Our knowledge of the internal structure and evolution of stars is mostly based on model computations,
which take into account and combine outcomes of almost all branches of modern physics.
To test how well our stellar models represent real stars,
very accurate observational constraints, such as mass, luminosity, radius, chemical composition, and rotational properties, 
are required. These constraints are obtained from observation of the outer regions, which 
depend on the physical conditions in the deep interior.

Time is one of the essential components of our universe.
Understanding the evolution of the far and near Universe   
depends on how successful we are in determination of the ages of 
astrophysical objects. In this respect, every type of age determination
is crucial for completing
the picture.
In this context, we consider  the observed detached eclipsing binaries (DEBs) that are members of open clusters, 
to obtain some basic information pertaining  to
stars. 

In the literature, thirteen well observed DEBs in twelve open clusters are found.
Among these systems, V818 Tau in Hyades has already been investigated in detail by the author and his colleagues (Y{\i}ld{\i}z {\it et al.} 2006) and is therefore is not included here.
Thus, twelve DEBs in eleven open clusters are studied, namely, V453 Cyg in NGC 6871, V1229 Tau (HD 23642) in Pleiades, V578 Mon in NGC 2244,
V906 Sco in NGC 6475, V497 Cep in NGC 7160, V381 Car (HD 92024) in NGC 3293, V392 Car in NGC 2516, V1034 Sco in NGC 6231, DW Car in Cr 228, GV Car in NGC 3532, V615 Per and V618 Per in NGC 869.
%These stars are very 'useful' with their best known age.  
The masses of the binaries vary between 1.54 M$_\odot$ and 16.84 M$_\odot$; therefore, this study essentially 
deals with the structure and evolution of the early-type stars. 
%{ Binaries with late-type components will be subject of another paper.} 

The key point for stellar evolution is the increase in mean molecular weight, 
%(or, decrease in number density of particles), 
which causes contraction in nuclear core. The temperature of the core is increased by the heat released 
as a result of contraction, and hence nuclear reactions are intensified. This is the basic mechanism 
that controls the evolution and determines the time dependence of stellar structure. 
{ By constructing rotating (for the rapidly rotating components of DW Car) and non-rotating }
evolutionary models for the component stars in the DEBs,
we obtain information about the age and chemical composition of the binaries and hence about their clusters. 
The information about age and chemical composition is deduced by fitting accurate values of luminosities and radii of 
component stars, if possible. We also obtain some simple expressions
based on the slope of mass--radius and mass--luminosity relations for the binaries
%in terms of ratio of radii and luminosities of component stars 
in order to find the age of binary systems easily. 

%burada
 
The ages of stars are derived by means of various astrophysical techniques
including 
(i) observed surface properties (see, for example, Kharchenko $et$ $al$., 2005; Y{\i}ld{\i}z $et$ $al.$, 2006); 
(ii) seismic properties (in particular for the stars with the solar-like oscillations, see for example, Eggenberger $et$ $al.$, 2008; Y{\i}ld{\i}z 2007, 2008);
%(iii) apsidal motion (see for example, Bak{\i}\c{s} $et$ $al$., 2008),
(iii) rotational properties (Mamajek \& Hillenbrand, 2008);             % Mamajek, Eric E.; Hillenbrand, Lynne A. 2008ApJ...687.1264
and (iv) radioactive elements (del Peloso $et$ $al.$, 2006).
The most widely used technique is based on the observed surface properties, which is also the method of the present study.

The Sun, with its precise seismic and non-seismic constraints, is the first target to model for an upgraded code describing 
stellar interiors.
The agreement between the solar models and observed constrains gives an idea of the quality of the code. 
{ One can obtain  the initial hydrogen ($X_{\rm o}$) and heavy element ($Z_{\rm o}$) abundances and 
the mixing-length parameter of the Sun from the calibration of solar models}, and use them 
when these quantities are not constrained for any star. In Y{\i}ld{\i}z (2008), $X_{\rm o}$, $Z_{\rm o}$ and $\alpha$ are found as 
0.70975, 0.016 and 1.82, respectively, for the solar model with heavy element mixture of Asplund,  Grevesse \& Sauval (2005) and 
with an age of 4.6 Gyr. The details of the code used in the model computations can also be found in 
{ Y{\i}ld{\i}z \& K{\i}z{\i}lo\u{g}lu (1997) and Y{\i}ld{\i}z (2007, 2008)}.

The remainder of this paper is organized as follows. In Section 2, we 
present the observational fundamental properties of the DEBs in open clusters available in the literature. 
The  models of their components are given in Section 3, { and the methods of age determination based on
the slopes of mass--radius and mass--luminosity relations }
%expressions derived from the models for age prediction 
are presented and discussed in Section 4. 
{ Error analysis for age is presented in Section 5.
Finally, we list some  
concluding remarks in Section 6.
}

\section[]{Observational Data of the Binaries in Open Clusters}
\begin{table*}
\caption{
 The fundamental properties of the double-lined eclipsing binaries, which are members of open clusters.
}
\label{ta1}
$%     $$
%\begin{array}{p{0.15\linewidth}cccccccp{0.15\linewidth}c}
%\begin{array}{p{0.13\linewidth}ccccrrrp{0.08\linewidth}l}
\begin{array}{p{0.13\linewidth}ccccllp{0.08\linewidth}l}
%\begin{array}{p{0.13\linewidth}cccclllll}
\hline 
           \noalign{\smallskip}

 %Star          &    $M/M_\odot$ & $R/R_\odot$  & $\logTe$      & $\logL$   &$\log(age(y))$ & $\log(age(y))$ & Cluster & Ref.Mv(En parlak)   B-V      U-B    kume adi    yildiz adi
 Star    & M/M_\odot & R/R_\odot  & \log(Te(K))      & \log(L/LR_\odot)   &t_{\rm W}(y)     & t_{\rm Kh}(y) &  Cluster & Ref. \\
\hline

V615 Per A& 4.075\pm 0.055 & 2.291\pm 0.141   &4.176 \pm 0.033 & 2.370\pm 0.080    & 1.172 \times 10^7   & 1.91 \times 10^7   & NGC869    & ~1 ^a\\
V615 Per B& 3.179\pm 0.051 & 1.903\pm 0.094   &4.041 \pm 0.045 & 1.820\pm 0.100    &           &        &         \\

V618 Per A& 2.332\pm 0.031 & 1.636\pm 0.069   &4.041 \pm 0.033 & 1.510\pm 0.140    & 1.172 \times 10^7   & 1.91 \times 10^7   &NGC869     & ~1 ^a\\
V618 Per B& 1.558\pm 0.025 & 1.318\pm 0.069   &3.903 \pm 0.045 & 0.770\pm 0.080    &           &        &          \\

V453 Cyg A& 14.36\pm 0.20 & 8.551\pm 0.055   &4.424 \pm 0.019 & 4.690\pm 0.210      & 9.08 \times 10^6    & 9.77 \times 10^6   &NGC6871 & ~2 ^a   \\
V453 Cyg B& 11.11\pm 0.13 & 5.489\pm 0.063   &4.406 \pm 0.031 & 4.240\pm 0.280      &           &        &  \\

V1229 Tau A& 2.193\pm 0.022 & 1.831\pm 0.029   &3.989 \pm 0.026 & 1.437\pm 0.047    & 1.5 \times 10^8     & 1.35 \times 10^8   &Pleiades & ~3~  \\
V1229 Tau B& 1.550\pm 0.018 & 1.548\pm 0.044   &3.880 \pm 0.053 & 0.858\pm 0.095    &           &        &\\

V578 Mon A& 14.54\pm 0.08 & 5.23\pm 0.06   &4.477 \pm 0.017 & 4.291\pm 0.188        & 7.870 \times 10^6   & 5.01 \times 10^6   &NGC2244    & ~4~  \\
V578 Mon B& 10.29\pm 0.06 & 4.32\pm 0.07   &4.421 \pm 0.015 & 3.901\pm 0.200        &           &        &\\

V906 Sco A& 3.253\pm 0.069 & 3.515\pm 0.039   &4.029 \pm 0.047 & 2.162\pm 0.082    & 2.400 \times 10^8   & 1.66 \times 10^8   & NGC6475& ~5~  \\
V906 Sco B& 3.378\pm 0.071 & 4.521\pm 0.035   &4.017 \pm 0.048 & 2.330\pm 0.084    &           &        &\\

V497 Cep A& 6.89\pm 0.46 & 3.69\pm 0.03   &4.290 \pm 0.021 & 3.265\pm 0.144        & 1.896 \times 10^7   & 4.57 \times 10^7   & NGC7160& ~6  ^a \\
V497 Cep B& 5.39\pm 0.40 & 2.92\pm 0.03   &4.249 \pm 0.023 & 2.860\pm 0.152        &           &        & \\

V381 Car A& 15.0\pm 3.0 & 8.4\pm 0.8   &4.406 \pm 0.020 & 4.419\pm 0.121            & 1.032 \times 10^7   & 8.71 \times 10^6   & NGC3293& ~7~ \\
V381 Car B& 3.0\pm 0.5 & 2.1\pm 0.4   &4.096 \pm 0.080 & 1.980\pm 0.290            &           &        &  \\

V392 Car A& 1.900\pm 0.02 & 1.625\pm 0.030   &3.947 \pm 0.023 & 1.162\pm 0.055     & 1.123 \times 10^8   & 1.20 \times 10^8   & NGC2516& ~8  \\
V392 Car B& 1.853\pm 0.02 & 1.600\pm 0.030   &3.937 \pm 0.023 & 1.109\pm 0.056     &           &        & \\

V1034 Sco A& 16.84\pm 0.48 &  7.45\pm 0.07 & 4.471 \pm 0.015&  4.580\pm 0.200       &  6.967 \times 10^6   & 6.46 \times 10^6  &NGC6231 & ~9  \\
V1034 Sco B& 9.38\pm 0.27 &  4.18\pm 0.04 & 4.370 \pm 0.033&  3.672\pm 0.210       &            &        & \\

DW Car A& 11.34\pm 0.12 &  4.558\pm 0.045 & 4.446 \pm 0.016&  4.055\pm 0.063     &  6.761 \times 10^6   & 4.79 \times 10^6  &Cr228   & 10  \\
DW Car B& 10.63\pm 0.14 &  4.297\pm 0.055 & 4.423 \pm 0.016&  3.915\pm 0.067     &            &        & \\

GV Car A& 2.51\pm 0.03 &  2.57\pm 0.05 & 4.009 \pm 0.030&  1.807\pm 0.219       &  3.105 \times 10^8   & 2.82 \times 10^8  &NGC3532 & 11  \\
GV Car B& 1.54\pm 0.02 &  1.43\pm 0.06 & 3.903 \pm 0.045&  0.875\pm 0.300       &            &        &     &   \\
            \noalign{\smallskip}
            \hline
\end{array}
$
{
$1$) Southworth {\it et al.} (2004a), $2$) Southworth {\it et al.} (2004b), $3$) Southworth {\it et al.} (2005), $4$) Hensberge {\it et al.} (2000), $5$) Alencar {\it et al.} (1997), $6$) Yakut {\it et al.} (2003), $7$) Freyhammer {\it et al.} (2005), $8$) Debernardi \& North (2001), $9$) Bouzid {\it et al.} (2005), $10$) Southworth \&Clausen (2007), $11$) Southworth \&Clausen (2006)\\
$^a$) The values given for the luminosities, radii and effective temperatures do not obey the luminosity equation $L=4\pi R^2 \sigma T_{\rm e}^4$ (see Section 3 and Table 3 ).
}
\end{table*}

   \begin{figure}
%\centerline{\psfig{figure=/home/yildiz/evol/BinClu/L.vs.Teff.ps,width=260bp,height=310bp}}
%\centerline{\psfig{figure=L.vs.Teff.ps,width=260bp,height=310bp}}
\centerline{\psfig{figure=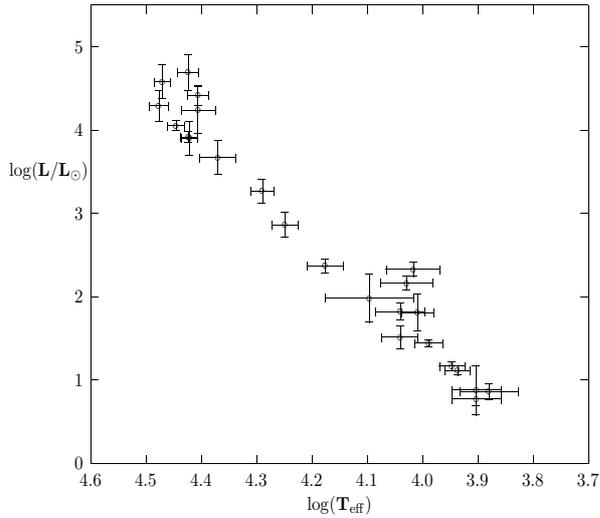,width=180bp,height=210bp}}
      \caption{Observational HR diagram for the DEBs in open clusters.
%($\odot$), $\alpha$ Cen A (filled circle) and B (star), 
}
              {\label{f1.1}}
   \end{figure}
{ Mass transfer obscures} the dependence of the fundamental properties of component stars 
on secondary causes,
such as, time, chemical composition, rotation, and magnetic field. Therefore, semi-detached and { contact} binaries are beyond the scope of the present study.
%in order to obtain information about the structure and evolution of $normal$ stars. 
Accordingly, DEBs, member of open clusters, are studied.  
%In order to obtain information about time dependence of fundamental properties (luminosity 
%and radius) of stars, we study detached binaries. Semi-detached and close binaries are 
%not considered, because
%mass transfer process obscure the variation of the fundamental properties of their components with time.  
We have found { in the literature studies} of 
twelve DEBs in eleven open clusters. 
The observed properties of their component stars are listed in Table 1.
The sixth and the seventh columns list the ages of the clusters given in WEBDA database \\
(http://obswww.unige.ch) and Kharchenko $et$ $al$. (2005), respectively.  
For some of the systems, $L=4\pi R^2 \sigma T_{\rm eff}^4$ is not satisfied (see Section 3 and Table 3).
%According to Southworth (private communication), this arises because the same distance for the two components of an DEB
%is enforced in the light-curve analysis.
%These stars are plotted in HR diagram in Fig. 1.
Fig. 1 shows the observed Hertzsprung-Russell diagram (HRD) for the component stars of these eclipsing binaries. 
All the stars are either main-sequence (MS) stars or very close to the MS.
The mass interval of the stars ranges from 1.54 (GV Car B) to 16.84 M$_\odot$ (V1034 Sco A) and it covers a large part of the MS in the HRD from O-type to F-type stars.
The ages of clusters vary between 5 My and 310 My. 
In the upper part of the MS in Fig. 1, from left to right, the ages of stars increase as theoretically expected. 
%in good agreement with what we theoretically expect. 
This result shows that the time dependence of effective temperature is more significant than any other second order dependence 
(e.g., chemical composition, rotation),
at least for these stars.

{
The eleven clusters hosting the studied eclipsing binaries are at distances ranging from 150 (Pleiades) to 2327 pc (NGC3293).
Number of stars in the clusters observed by UBV photometry is good enough to estimate the ages of the clusters by isochrone fitting.
%The ages of the clusters are taken from the WEBDA data base (http://www.univie.ac.at/webda) and Kharchenko {\it et al.} (2005).
%WEBDA compiles the ages from the literature. 
The WEBDA ages 
taken from different studies vary. Kharchenko {\it et al.} (2005)
essentially use the isochrones of Girardi {\it et al.} (2002) for age determination. 
%which also include post-MS stages.
}

%\section{Results and Discussions}
%The luminosities of the stars listed in Table 1 are plotted with respect to stellar mass in Fig. 2. The solid line in Fig. 2 represents a line fit. 

{ These binaries have attracted attention in recent years. None of them is among the well known binaries 
in Andersen (1991).
However, six of them (V453 Cyg, V578 Mon, V906 Sco, V392 Car, V1034 Sco and DW Car) are listed in the recent 
compilation by Torres et al. (2010). Here we shall compare the basic properties of the stars in Table 1 
with that of given in 
Andersen (1991).

{ The slope of the mass--luminosity relation derived using the  data in Table 1 is 3.89. This means that 
\begin{equation} 
L\propto M^{3.89}.
\end{equation} 
%log L/log M=3.893116
For the same mass range,  we find $L\propto M^{4.02}$ for  the DEBs with the most precise data 
on the fundamental properties (Andersen 1991). 
%L = M^3.97808
These two values for the power of $M$ are very close to each other.   
%In Fig. 3, the radii of the stars are plotted with respect to stellar mass. The solid line shows a line fit and its slope is 0.70: $R\propto M^{0.70}$.
 The slope of the mass--radius relation of the stars in Table 1 is found as 0.704. 
That is to say $R\propto M^{0.704}$.}
%R = M^0.813364
For the well-known eclipsing binaries (Andersen 1991), we find that $R\propto M^{0.82}$.
The slopes we find for the mass--radius and the mass--luminosity relations for binaries given in Table 1 are 
slightly less than the slopes found from the well known binaries.  
%R\propto M^0.813364
The reason may be that 
many stars in the $log(M/M_\odot)$ interval [0.1,0.4] in Andersen (1991) have
larger radii than their terminal-age MS (TAMS ) radius. 
If we take the mass range as [2.5 M$_\odot$, 16.84 M$_\odot$] rather than [1.54 M$_\odot$, 16.84 M$_\odot$], 
the slopes of the mass--luminosity and the mass--radius relations for the well known binaries are  
found as { 3.94 and 0.723}, respectively. These values are very close to the values we find for the eclipsing 
binaries in the open clusters.}

%For the masses greater than $log(M/M_\odot)>0.4$, i.e. $M> 2.5 M_\odot$, the slopes computed for the stars in Andersen (1991) are the 
%give the same value
%for { the slope of} the mass--radius relation as the eclipsing binaries in open clusters. 

%We wonder if the fundamental properties of the stars given in Table 1 show any time dependence.
%For the time dependency of $L$, we plot $\log(L/L_\odot)-3.89\log(M/M_\odot)$ with respect to the age of the cluster in Fig. 2.
%There seems to exist a relation between $\log(L/L_\odot)-3.89\log(M/M_\odot)$ and time. 
%A similar situation is also valid for the mass-radius relation. In Fig 3., $\log(R/R_\odot)-0.704\log(M/M_\odot)$ is plotted with respect to $\log(t(y))$.
%
%
%In Fig. 4, ages of the stars are plotted with respect to stellar mass. Components of an eclipsing binary are connected with lines. In order to show the evolutionary stage of the stars in MS band,
%the Kelvin-Hemholtz (lower thin solid line) and nuclear time scales (upper thick solid line) are also plotted in Fig. 4, which schematically show the evolutionary stages of each component.  
%Also shown in Fig. 4 is half of the nuclear time scale (dotted line).
%
%The Sun spends 11 Gy on the main-sequence (Sackmann, I.-J., Boothroyd, A.I.; Kraemer, K.E. 1993, ApJ, 418, 457)

\section{Ages of the Clusters from Models of the Component Stars in the Eclipsing Binaries}

%The difference between the masses of the components of  V906 Sco is not very large, but the primary component is a TAMS
%star and therefore very appropriate for age determination. 
Thw age of a binary system can be  found from the solution of two equations
based on the coevality of the component stars 
with two unknowns. 
%are listed in Table 3. These two equations are based on the coevality of the component stars.
One of the equations is written for the difference between the 
ages of the two components of the binary based on observed luminosities: 
\begin{equation}
\Delta t_{\rm L}= t_{\rm LA}-t_{\rm LB},
\end{equation}
where $t_{\rm LA}$ and $t_{\rm LB}$ are ages found from luminosities of primary and secondary components,
respectively.
The other equation basewd on the radii:
\begin{equation}
\Delta t_{\rm R}= t_{\rm RA}-t_{\rm RB},
\end{equation}
where $t_{\rm RA}$ and $t_{\rm RB}$ are ages from the radii of primary and secondary components,
respectively.
From the assumption of coevality, $\Delta t_{\rm L}=\Delta t_{\rm R}=0$. 
Using numerical derivatives of $\Delta t_{\rm L}$ and $\Delta t_{\rm R}$ with X and Z,
we solve these equations and find X and Z. In general, 
{ we find that different
combinations of X and Z for a given binary give similar results; thus,}
%different reference models give different combinations of X and Z; hence, 
the solution is not unique. 
If it is not possible to obtain an acceptable solution to Eqs. (2) and (3) for a binary system, then we either use one of these equations or
fit the model of one of its components to observed values.

{ In the following subsections, interior models of the component stars in Table 1 constructed using the stellar evolution code described in  Y{\i}ld{\i}z \& K{\i}z{\i}lo\u{g}lu (1997) and Y{\i}ld{\i}z (2007, 2008) are presented. From these models, 
the ages and chemical composition of the binaries are obtained if possible.
}
\subsection{On the Fundamental Properties of Components of V615 Per and V618 Per}
%2004MNRAS.349..547S Southworth, J.; Maxted, P. F. L.; Smalley, B.
In this subsection, we consider the binary systems V615 Per and V618 Per in more detail.
Their components give very different ages. The reason for this difference shoul be investigated.
%
%Three systems, namely, V615 Per, V578 Mon and V906 Sco, 
%have very different { slope for} the mass--luminosity relation ($l_m=\log(L_{\rm A}/L_{\rm B})/\log(M_{\rm A}/M_{\rm B})$)
%from the others. The reason for the discrepancy for V906 Sco is 
%obvious; its primary (massive) component is evolved and therefore it is brighter and 
%also larger than its normal MS counterparts. For V578 Mon the situation is more complicated. 
%
For V615 Per, we compare its components with those of V618 Per and find that the luminosity 
of V615 Per B is lower than is exppected for its mass. If we assume that the power of mass 
is 4 for {the slope of the} mass--luminosity relation, then $\log (L_B/L_\odot)=2.02$, which is larger 
than the value found from observations with by amount of 0.2. Since { the slope of} the mass--radius relation for 
the components of V615 Per and V618 Per seems reasonable, we deduce that the effective 
temperature of V615 Per B should be larger than the value found by Southworth, Maxted \& Smalley (2004a). 
As a matter of fact, the results they give for the effective temperatures of V615 Per B 
($M_B= 3.18\pm 0.05$ M$_\odot$) and V618 Per A ($M_A= 2.33 \pm 0.03$ M$_\odot$) are the 
same, despite the very different masses. 
%Then, assuming the underestimation of luminosity 
%of V615 Per B is due to underestimation of its effective temperature, we find that 
%$log(T_e)$ of V615 Per B must be 4.09 (see below and Table 3). 

%The reasonable models for the components of V618 Per are obtained with Z=0.01 and X=0.735.
%These models are plotted in the HRD in Fig. 6. The observed values are also indicated.
%Also shown in Fig. 6 is the position of the models at the ages of the cluster given in WEBDA and by
%Kharchenko et al. For V618 Per A the models share the same place in HRD. For V618 Per B, 
%the age=1.91 \times 10^7 yrs (Kharchenko et al. ) is more suitable than 1.0 \times 10^7 yrs (WEBDA).
%At 1.0 \times 10^7 V618 Per B is in PMS stage. 

Reasonable models for the components of V618 Per are obtained with Z=0.01 and X=0.735. In the HRD,
the positions of the models at the ages of the cluster given in WEBDA and by
Kharchenko {\it et al.} are the same for V618 Per A. For V618 Per B,
the age=19.1 My (Kharchenko {\it et al.} 2005) is more suitable than 11.7 My (WEBDA).
At 11.7 My, V618 Per B is in the pre-MS stage. The ages and the chemical compositions we find are listed 
in Table 2.

The accurate values of the components of V615 and V618 Per given in Southworth (2004a) do not exactly obey $L=4\pi R^2 \sigma T_{\rm eff}^4$.
%For V618 Per A and B, it seems that the luminosities are missprinted. 
According to Southworth (private communication), this arises because the same distance for the two components of a DEB
is enforced in the light-curve analysis.
For the system V615 Per, however, the problem is with V615 Per B, particularly
with its effective temperature. 
For its $\log (L/L_\odot) =1.82$ and $ R/R_\odot=1.903$ values, its effective temperature
should be 12000 K rather than 11000 K. 
%However, the given luminosity is significantly low compared to its mass.
%For convenience, $\log (L/L_\odot )=1.960$ and then $T_{\rm eff}=13000$.
The suggested values are given in bold style in Table 3. 

{ For age determination of these systems, we use observational radii of the component stars 
as constraints for their interior models, because of the complicated luminosities and effective temperatures. 
The age of these systems is roughly estimated as 20 My. 
V615 Per B seems to be a MS star at this age. 
%The ages for V615 Per and V618 Per found from ratio of radii are in very good agreement 
}

\begin{table*}
\caption{
 The chemical composition and age of the binary systems derived from the calibration of models.
 The ages we find are listed in the fourth column. For comparison, the ages of the clusters given
 in the WEBDA database and by Kharchenko {\it et al.} (2005) are also presented. The seventh column reports
 the MS life time of the primary components of the binary systems. In the eighth column,
 the age found from the ratio of radii of the component stars is shown (see Section 4).
{ The ages found from the fitting formula given in Eqs. 6-9 are listed in ninth column}
}
\label{ta1}
$%     
%\begin{array}{p{0.15\linewidth}cccccccp{0.15\linewidth}c}
\begin{array}{p{0.11\linewidth}llclllllp{0.1\linewidth}}
\hline 
            \noalign{\smallskip}
 %Star          &    $M/M_\odot$ & $R/R_\odot$  & $\logTe$      & $\logL$   &$\log(age(y))$ & $\log(age(y))$ & Cluster & Ref.Mv(En parlak)   B-V      U-B    kume adi    yildiz adi
%Star          &     X_0   & Z_0    & t &      t(W)     & t(K) & t_{\rm MSA}  & Cluster \\
% Star          &     X$_0$   & Z$_0$    & t(y) &      t$_{\rm W}$(y)     & t$_{\rm Kh}$(y) & $t$_{\rm MSA}$ & $t_{\rm rm}$& {\rm Cluster} \\
 Star    &     X   & Z    & t(y) &      t_{\rm W}(y)     & t_{\rm Kh}(y) & t_{\rm MSA} & t_{\rm rm}& t_{\rm rmf} & {\rm Cluster} \\
% Star          &     X_0   & Z_0    & t(y) &      t_{\rm W}(y)     & t_{\rm Kh}(y) & t_{\rm MSA} & t_{\rm rm}& {\rm Cluster} \\
\hline
V615 Per &    0.7350 & 0.0100 & 2 \times 10^7   &1.172 \times 10^7   & 1.91 \times 10^7  & 1.40 \times 10^8    & 2.92  \times 10^7 & 2.71 \times 10^7 &{\rm NGC~869}\\
V618 Per &    0.7350 & 0.0100 & 2 \times 10^7   & 1.172 \times 10^7   & 1.91 \times 10^7 & 5.98 \times 10^8    & ~--&  ~ -- & {\rm NGC~869}\\
V453 Cyg &   0.7046 & 0.0117 & 1.05 \times 10^7   & 9.08 \times 10^6    & 9.77 \times 10^6 & 1.18 \times 10^7  & 9.76 \times 10^6 &  9.60 \times 10^6 &{\rm NGC~6871}\\
V1229 Tau&    0.6946 & 0.0158 & 1.93 \times 10^8  & 1.5 \times 10^8     & 1.35 \times 10^8 & 6.88 \times 10^8  & 1.24 \times 10^8 &  1.44 \times 10^8 &{\rm Pleiades}\\
V578 Mon &   0.7098 & 0.0160 &  3.50 \times 10^6     & 7.870 \times 10^6   & 5.01 \times 10^6 & 1.29 \times 10^7 & ~-- &  ~ -- &{\rm NGC~2244}\\
V906 Sco &    0.7200 & 0.0170 & 2.52 \times 10^8  & 2.400 \times 10^8   & 1.66 \times 10^8 & 2.53 \times 10^8   & 2.53 \times 10^8 &  2.53 \times 10^8 &{\rm NGC~6475}\\
V497 Cep &    0.7440 & 0.0125 & 2.30 \times 10^7  & 1.896 \times 10^7   & 4.57 \times 10^7 & 4.95 \times 10^7   &  2.17 \times 10^7 &  2.03 \times 10^7 &{\rm NGC~7160}\\
V381 Car &    0.7380 & 0.0258 & 1.05 \times 10^7 & 1.032 \times 10^7   & 8.71 \times 10^6  & 1.37 \times 10^7  & 1.05 \times 10^7 &  0.43 \times 10^7 &{\rm NGC~3293}\\
V392 Car &    0.6837 & 0.0178 & 1.23 \times 10^8  & 1.123 \times 10^8   & 1.20 \times 10^8 & 1.04 \times 10^9  & 1.41 \times 10^8 &  3.01 \times 10^8 &{\rm NGC~2516}\\
V1034 Sco&    0.6961 & 0.0226 & 5.51 \times 10^6  &  6.967 \times 10^6   & 6.46 \times 10^6 & 1.02 \times 10^6 & 5.36 \times 10^6 &  4.21 \times 10^6 &{\rm NGC~6231}\\
DW Car   &    0.6934 & 0.0106 & 5.22 \times 10^6  &  6.761 \times 10^6   & 4.79 \times 10^6 & 1.56 \times 10^7 & 4.88 \times 10^6 &  5.72 \times 10^6 &{\rm Cr~228}\\
" (Rotat.)&    0.6951 & 0.0078 & 5.29 \times 10^6  &  6.761 \times 10^6   & 4.79 \times 10^6 & 1.52 \times 10^7  & 4.75 \times 10^6 &  5.75 \times 10^6 &{\rm Cr~228} \\
GV Car    &    0.7046 & 0.0104 & 3.22 \times 10^8  &  3.105 \times 10^8   & 2.82 \times 10^8 & 4.22 \times 10^8 & 3.21 \times 10^8 &  3.04 \times 10^8 &{\rm NGC~3532}\\
            \noalign{\smallskip}
            \hline
\end{array}
$

\end{table*}

\begin{table*}
\caption{
 The modifications in the fundamental properties of the double lined eclipsing binaries due to inconsistent results given
in the literature are presented in bold style
}
\label{ta1}
$%     $$
%\begin{array}{p{0.15\linewidth}cccccccp{0.15\linewidth}c}
\begin{array}{p{0.16\linewidth}ccccrrrp{0.09\linewidth}r}
\hline
            \noalign{\smallskip}

 %Star          &    $M/M_\odot$ & $R/R_\odot$  & $\logTe$      & $\logL$   &$\log(age(y))$ & $\log(age(y))$ & Cluster & Ref.Mv(En parlak)   B-V      U-B    kume adi    yildiz adi
 Star          &    M/M_\odot & R/R_\odot  & \log T_{\rm eff}({\rm K})      & \log(L/L_\odot)   \\
\hline

V615 Per  A    &    4.075\pm 0.055 & 2.291\pm 0.141   &4.176 \pm 0.033 & 2.370\pm 0.080    \\
V615 Per  B    &    3.179\pm 0.051 & 1.903\pm 0.094   &{\bf 4.079} \pm 0.045 & 1.820\pm 0.100    \\

%V615 Per  A    &    4.075\pm 0.055 & 2.291\pm 0.141   &4.176 \pm 0.033 & 2.370\pm 0.080    \\
%V615 Per  B    &    3.179\pm 0.051 & 1.903\pm 0.094   &{\bf 4.114} \pm 0.045 & {\bf 1.960}\pm 0.100    \\

V618 Per  A    &    2.332\pm 0.031 & 1.636\pm 0.069   &4.041 \pm 0.033 & {\bf 1.546}\pm 0.140    \\
V618 Per  B    &    1.558\pm 0.025 & 1.318\pm 0.069   &3.903 \pm 0.045 & {\bf 0.805}\pm 0.080    \\

V453 Cyg  A    &   14.36\pm 0.20 & 8.551\pm 0.055   &4.424 \pm 0.019 & {\bf 4.513}\pm 0.210      \\
V453 Cyg  B    &   11.11\pm 0.13 & 5.489\pm 0.063   &4.406 \pm 0.031 & {\bf 4.056}\pm 0.280      \\

V497 Cep  A    &    6.89\pm 0.46 & 3.69\pm 0.03   &4.290 \pm 0.021 & {\bf 3.245}\pm 0.144        \\
V497 Cep  B    &    5.39\pm 0.40 & 2.92\pm 0.03   &4.249 \pm 0.023 & {\bf 2.877}\pm 0.152        \\
            \noalign{\smallskip}
            \hline
\end{array}
$
\end{table*}

\subsection{On the Fundamental Properties of Components of V906 Sco}
%Alencar, S. H. P.; Vaz, L. P. R.; Helt, B. E. 1997A&A...326..709A
The difference between the masses of the components of  V906 Sco is not very large but the primary component is 
a TAMS
star and therefore very appropriate for age determination. The ages are found from the solution of Eqs. (2) and (3),
based on the coevality of the component stars.
We solve these equations and find X and Z. Using different reference models, we obtain very different combinations of
X and Z, hence the solution is not unique. 
However, the ages found from luminosities and radii of V906 Sco A and B are very close to each other.
For the three cases given in Table 4, the age found from $R_{\rm B}$ (the last column) 
is slightly less than the others. The mean value of ages is 252 My. Without $t_{\rm RB}$ it is 256 My. 

\begin{table*}
\caption{
 Ages derived from fitting luminosities ($t_{\rm L}$) and radii ($t_{\rm R}$) of models for V906 Sco A and B to the observed values, for different
combinations of hydrogen and heavy element abundances. While the  mean of all values is 252 My, the mean value of $t_{\rm LA}$, $t_{\rm LB}$ and $t_{\rm RA}$ 
is 256 My.
}
\label{ta1}
$%     $$
\begin{array}{p{0.15\linewidth}ccccccp{0.15\linewidth}c}
\hline 
            \noalign{\smallskip}
  X & Z  & t_{\rm LA}(My)  & t_{\rm LB}(My) & t_{\rm RA}(My)  & t_{\rm RB}(My) \\
\hline
0.720   &0.0170 & 256  & 256  & 253 & 235    \\
0.729   &0.0156 & 256  & 259  & 258 & 240    \\
0.737   &0.0142 & 257  & 260  & 253 & 244    \\

            \noalign{\smallskip}
            \hline
\end{array}
$
\end{table*}

\subsection{V578 Mon}
This system has the lowest values for 
{ slopes of the} mass--luminosity and the mass--radius relations
among the systems with the early-type components:\\
$l_m=\log(L_{\rm A}/L_{\rm B})/\log(M_{\rm A}/M_{\rm B})=2.6$, \\
$r_m=\log(R_{\rm A}/R_{\rm B})/\log(M_{\rm A}/M_{\rm B})=0.55$.

This means that either the observational values of $L_{\rm A}$ and $R_{\rm A}$  are very low or 
the the observational values of $L_{\rm B}$ and $R_{\rm B}$ are very high, compared to the values expected for their masses.
The influence of this point on the age determination of the system is crucial.

Abundances of some heavy elements (including C, N and O) are found from disentangled spectra by Pavlovski and Hensberge (2005).
The oxygen is underabundant by about -0.10 dex. If we take the solar $Z_s=0.0122$ 
{ (surface heavy element abundance)}, 
then $Z_0$  of NGC 2244 is about 0.01.

From the models of the component stars with solar composition ($X_{\rm o}=0.70975, ~ Z_{\rm o}=0.016$), we find the agreement time for the quantities $L_{\rm A}$,
$L_{\rm B}$, $R_{\rm A}$, and $R_{\rm B}$. While $t_{L\rm A}$ and  $t_{R\rm A}$ are about 3.5 My, $t_{L\rm B}$  and $t_{R\rm B}$ are about three and two times
larger than 3.5 My, respectively. 
The reason for the small radius of V578 Mon A or large radius of V578 Mon B may be the accuracy of the data. However, if this is not the case then this system
has some interesting features which are beyond the standard models, for example, existence of rapidly rotating core
for V578 Mon A (Y{\i}ld{\i}z 2003, 2005) or  rotational mixing (Meynet \& Maeder 2000). Within the standard approach, the present data implies that V578 Mon B is more $evolved$
than V578 Mon A. The compatible radius for V578 Mon A based to its mass is about 6 $R_\odot$, which 
is 15 \% larger than the observed value. Assuming the same effective temperature as given by Hensberge {\it et al.} (2000), the increase in luminosity of V578 Mon A due to 
the increase in its radius is about 30 \%. For this radius and luminosity, the age of the system is  found as 6 My from the models of V578 Mon A and B with Z=0.01 and X=0.710.
%However, the problem may be due to the observed radius of V578 Mon B. In that case the

Hensberge {\it et al.} (2000) shows that the position of V578 Mon A in HRD corresponds to 11 $M_\odot$.  
This may be a manifestation of the rapidly rotating cores of early type stars.

\subsection{V392 Car}
%Debernardi, Y.; North, P. 2001A\&A...374..204 fundamental data
%CP yildizi mi? Hayir
The masses of the components of V392 Car are very close to each other: $M_{\rm A}=1.90\pm 0.02$ M$_\odot$, $M_{\rm B}=1.853\pm 0.02$ M$_\odot$ (Debernardi \& North 2001).
In Table 5, the ages found from the luminosities and radii for different combinations of $X$ and $Z$ are listed.
%{ The ages given in the last column of Table 5 is derived by using the ratio of radii rather than the 
%individual values of luminosities or radii. }
For such systems with similar components, small differences due to uncertainties in the observed quantities may lead very different results in model computations.
However, the observed values of the radii of the component stars are very accurate and provide an opportunity to 
determine  the age of the system very accurately. { The models with $X=0.6837$ and $Z=0.0178$ predict very 
close ages for V392 Car A and B (126 and 122 My, respectively). 
%From the ratio of radii of the component stars, we find 125 My for the age of the system. 
The adopted value for the age, 124 My, is the mean of these values. 
}
{ Debernardi \& North (2001) reported very similar metallicity for V392 Car, Z=0.018, using evolutionary tracks of Schaller {\it et al.} (1992).}

%CP yildizi mi? Hayir

\begin{table*}
\caption{
 Ages derived from fitting the luminosities ($t_{\rm L}$) and radii ($t_{\rm R}$) of models for V392 Car A and B to the observed values, for different
combinations of hydrogen and heavy element abundances %The ages given in the last column are found from the ratio of the radii.
}
\label{ta1}
$%     $$
\begin{array}{p{0.15\linewidth}ccccccp{0.15\linewidth}}
\hline 
            \noalign{\smallskip}
  X & Z  & t_{\rm LA}(My)  & t_{\rm LB}(My) & t_{\rm RA}(My)  & t_{\rm RB}(My) \\ %&t_{\rm RA/RB}(My)\\
\hline
%0.720   &0.0135 & 260  & 200  & 273 & 284    \\
%0.702   &0.0155 & 200  & 156  & 189 & 203    \\
%0.690   &0.0174 & 220  & 182  & 140 & 156    \\
%0.688   &0.0184 & 246  & 182  & 112 & 105    \\
%
0.7200  & 0.0135 & 260 &  200 &  273  &284 \\ % &170 \\
0.7020  & 0.0155 & 200 &  156 &  189  &203 \\ % &113 \\
0.6900 & 0.0174 & 220 &  182 &  140  &156 \\ % &103 \\
%0.6837 & 0.0178 & 157 &   98 &  123  &122  & 125 \\
%0.6825 & 0.0180 & 163 &  101 &  119  &116  &140 \\
0.6837 & 0.0178 & 157 &   98 &  126  &122 \\ % & 125 \\
0.6825 & 0.0180 & 163 &  101 &  119  &116 \\ % &140 \\
0.6880 & 0.0184 & 246 &  182 &  112  &105 \\ % &165 \\
0.6800 & 0.0184 & 165 &   99 &  111  &102 \\ % &188 \\

            \noalign{\smallskip}
            \hline
\end{array}
$
\end{table*}

\subsection{V1229 Tau (HD 23642)}

%Groenewegen, M. A. T., Decin, L., Salaris, M., De Cat, P. 2007, A\&A, 463, 579
%Munari, U., Dallaporta, S., Siviero, A., Soubiran, C., Fiorucci, M., Girard, P. 2004, A\&A, 418, 31
%Southworth, J., Maxted, P. F. L., Smalley, B. 2005, A\&A, 429, 645

The fundamental properties of components of V1229 Tau are determined from light and radial velocity curves by Munari {\it et al.} (2004), Southworth {\it et al.} (2005) and Groenewegen {\it et al.} (2007).
The results of these studies are very similar. 
Among the eclipsing binaries in open clusters considered, this system 
has the minimum value of \\
$r_m=\log(R_{\rm A}/R_{\rm B})/\log(M_{\rm A}/M_{\rm B})$.
%(Munari et al. (2004) Southworth et al. (2005) and Groenewegen et al. (2007)).

It is not possible to fit models of V1229 Tau A and B to the
observed properties simultaneously. 
Therefore, we find age of the system by considering V1229 Tau A and B separately.
%There are two possible cases: (i) The problem is with V1229 Tau B. When we fit radius model V1229 Tau A to the observed radius,
%model radius of V1229 Tau B with the same chemical composition is about ..\% is smaller than the 
%observed radius. For luminosity of V1229 Tau B, the difference is about ... \%.
%(ii) The problem is with V1229 Tau A.
%If we fit radius of V1229 Tau B  and find that model radius of V1229 Tau A with the same chemical composition
%is about ..\% greater than the observed radius.  

{ The age derived from the calibration of V1229 Tau A is 193 My for ($X,Z$) = (0.6946, 0.01575) and 222 My for ($X,Z$) = (0.7104, 0.014).
However, the age derived from the calibration of V1229 Tau B is
730 My for ($X,Z$)=(0.7046,0.01305). This value is about 
five times greater than the value given by Kharchenko $et$ $al$. 
In comparison to the age found for V1229 Tau B, the ages found for V1229 Tau A are close to the ages of 
Kharchenko $et$ $al$. }
{ Southworth {\it et al.} (2005) finds an age about 125-175 My, close to the results 
from the calibration of V1229 Tau A. } The age of V1229 Tau is listed as 193 My in Table 2.

\subsection{V453 Cyg}
%Southworth, J., Maxted, P. F. L., Smalley, B. 2004, MNRAS, 351, 1277
Recently, precise fundamental properties of V453 Cyg A and B have been presented by 
Southworth, Maxted \& Smalley (2004b).
However, there is a contradiction in their results. According to the radii and 
effective temperatures of the components, the 
luminosities { must be  lower than the published values}; in their Table 11, they give \\
$\log (L_{\rm A}/L_\odot)=4.69$ 
and $\log (L_{\rm B}/L_\odot)$=4.24 but we compute the luminosities from the radii and effective temperatures given in that table
and find that $\log (L_{\rm A}/L_\odot)=4.513$ and $\log (L_{\rm B}/L_\odot)$=4.056. The correction in $\log (L)$ is about 0.18;
about 50\% (see Table 3).

From the calibration of model radii of V453 Cyg A and B for X=0.700, we find the age of the system as 10.3 My and Z=0.0174.
For X=0.7046, however, the age and Z are computed as 10.5 My and 0.0117, respectively. 
Whereas the ages found with very different chemical compositions are very close, 
the model luminosities with the latter values are in perfect agreement with the (corrected) values found from observations.
Southworth, Maxted \& Smalley (2004b) give the age of the system as 10 My. 
The ages determined for V453 Cyg are in good agreement with this age and the age of the cluster given by Kharchenko {\it et al.} (2005).
Such a good agreement is based on the fact that V453 Cyg A is very close to the TAMS.

Abundance analysis of V453 Cyg A and B is made by Pavlovski \& Southworth (2009). They found that N, O and Mg abundances for V453 Cyg A are
nearly solar and C, Si and Al are underabundant. Regarding the fact that O is the most abundant heavy element for $normal$ stars, we estimate that 
$Z \simeq 0.01$ from oxygen abundance (-0.08 dex) estimated by Pavlovski \& Southworth (2009). This value is in good agreement with the values predicted from the models (Z=0.0117).
Pavlovski \& Southworth  also re-derived the effective temperatures of V453 Cyg A and B: $T_{\rm effA}=27,900\pm 400$ K and $T_{\rm effB}=26,200\pm 500$ K. These values are 5 and 
3\% greater than the values given by Southworth, Maxted \& Smalley (2004b).

Pavlovski \& Southworth  forecast the helium abundance as 0.086 and 0.123 dex, depending upon the method for derivation of 
microturbulence velocity. { For these helium abundances, 
very similar ages are found: for 0.086 dex, Z=0.0113 and the mean age is 10 My; for 0.123 dex, 
Z=0.0163 and the mean age is 9 My.
%no simultaneous solution is found for V453 Cyg A and B. 
%The separate solutions with different values of Z for V453 Cyg A and B give that the age is  1.0 My.
}
%  OOKK

%From the calibration of radii 
%t8=1.03  it.B.V453Cyg.0.700.0.0174.FIT  logLA=4.478 logLB=4.0307    0.4473\\
%                                        logTA=4.4152 logTB=4.3997   0.0155\\
%                                        TA= 26014        TB= 25102\\
%
%t8=1.05  fit.B.V453Cyg.0.70458.0.0117.FIT logLA=4.5076 logLB=4.0667  0.4409\\
%                                        logTA=4.4233 logTB=4.4109    dlogT=0.0124\\
%                                        TA= 26503        TB= 25757\\
%
%gozlem logLA=4.512 logLB=4.0553           dlogL=0.45\\
%       logTA=4.424 logTB=4.406               0.018\\
%       TA= 26503        TB= 25757
%
%
%
%
%The luminosities of models V453 Cyg A and B at these times are very close to the lower limit of the observed 
%values. In that case, the effective  

\subsection{V497 Cep}
% Yakut et al. (2003)
The fundamental properties of this system were obtained by Yakut {\it et al.} (2003). Using their findings,
the best agreement between the models and the observational results from photometric and spectroscopic analysis is 
reached with X=0.744 and Z=0.0125. The derived age 23.0 My is the average of 
$t_{\rm LA}=t_{\rm RA}=$ 22.3 My,  $ t_{\rm LB}$=24.9 My,  and $t_{\rm RB}$=22.5 My.
Whereas the age we find is close to the age for the cluster given in WEBDA,  it is half of the 
value derived by Kharchenko {\it et al.} (2005).

The accurate values given in Yakut {\it et al.} (2003) do not exactly obey $L=4\pi R^2 \sigma T_{\rm e}^4$.
It is not easy to find the source of this inconsistency. However, it may be due to round off.
% OOKK

\subsection{V381 Car (HD 92024)}
%Freyhammer, L. M.; Hensberge, H.; Sterken, C.; Pavlovski, K.; Smette, A.; Ilijić, S.  2005A&A...429..631F
%Freyhammer, L. M.; Hensberge, H.; Sterken, C.; De Cat, P.; Aerts, C. 2006, Mem. S.A.It., 77, 334    beta cep 3 osc. freq.  (l=2,4)
V381 Car A is a $\beta$ Cep for which three oscillation frequencies are observed with l=2 and 4 (Freyhammer {\it et al.}, 2006). 
Despite the large uncertainty in the fundamental properties of its components (Freyhammer {\it et al.}, 2005), it is a normal system with its { slopes for the} mass--radius and mass--luminosity relations
given as  0.861 and 3.4894. 
From the fit of model radii and luminosities to the observed values, with X=0.738 and Z=0.0258, 
we find that $t_{\rm LA}=t_{\rm RA}=$ 10.8 My and  $t_{\rm RB}$=9.8 My. The age of the system
is found as 10.5 My, the average of these values. { This age is in agreement with the age range (10-13 My) estimated by 
Freyhammer {\it et al.} (2005). }

The chemical composition found from fitting interior models of V381 Car A and B is very different from that of the other
binaries in the sense that both of X and Z are greater than the solar values.
This may be a result of the over-estimated masses of V381 Car A and B.
\subsection{V1034 Sco}
%Bouzid, M. Y.; Sterken, C.; Pribulla, T.  2005A&A...437..769B
It is a normal system with its { slopes for the } mass--radius and mass--luminosity relations of 0.9876 and 3.5728,
respectively.
From the fit of model radii and luminosities to the observed values, with X=0.6961 and Z=0.02255, 
we find that $t_{\rm RA}=t_{\rm RB}=$ 5.64 My,  $ t_{\rm LA}$=5.72 My and $t_{\rm LB}$=5.05 My. 
The age of the system is computed as 5.51 My by taking average of these values. {  Bouzid {\it et al.} 
(2005) confirm that the age of the system is about 5 My, in good agreement with the findings of the present study.} 
\subsection{DW Car}
%Southworth, J.; Clausen, J. V., 2007A&A...461.1077S
At first sight, this system seems a normal system with {slopes for the } mass--radius and mass--luminosity relations 
of 0.9120 and 4.9858, respectively (Southworth \& Clausen, 2007).
According to the radii, the components of DW Car are unevolved MS stars and very close the zero-age MS (ZAMS).
However, the luminosity derived from the observation implies that DW Car A is an evolved MS star.
No simultaneous solution is available for the luminosities of DW Car A and B.
From their radii, the age of the system is calculated as 5.22 My. The non-rotating models giving this age 
are constructed with X=0.6934 and Z=0.01063. However, DW Car A and B are fast rotators; $v_{\rm eqA}=182\pm 3$ km s$^{-1}$ and  $v_{\rm eqB}=177\pm 3$ km s$^{-1}$.
Therefore, models rotating like a solid body are also constructed for them. The calibration of rotating models results in nearly the same age, 5.31 My.
The hydrogen and metal abundances for the rotating models are X=0.6951 and Z=0.0078, very close to that of non-rotating models. 
{ Southworth \& Clausen (2007) reported similar results: about 6 My, $Z\simeq 0.01$.}
   
%Comparison of radii ---> the components are unevolved MS stars.
%
%Comparison of luminosities ---> the primary component is  evolved MS star.

\subsection{GV Car}
This system consists of two metallic-line stars and it shows apsidal motion. From the calibration of models of the components,
with X=0.7046 and Z=0.0104,
we estimate that $t_{\rm RA}=$324 My, $t_{\rm RB}=$ 372 My,  $ t_{\rm LA}$= 295 My and $t_{\rm LB}$=302 My.
The age of the system is found as 322 My. 
Southworth \& Clausen (2006) give a very similar age for GV Car, 360 My.

%For ($X_0,Z_0$)=(0.7046,0.0104), tRA=8.51, tLA=8.47\\ 
%                                 tRB=8.57, tLB=8.48
 
\section{Mass and Time Dependence of the Mass--radius and the Mass--luminosity Relations }
\begin{table*}
\caption{
Luminosity and radius difference between ZAMS and TAMS for evolutionary tracks of V906 Sco A. 
$\Delta \log(L)=(L_{\rm TAMS}-L_{\rm ZAMS})/L_{\rm TAMS}$, $\Delta \log(R)=(R_{\rm TAMS}-R_{\rm ZAMS})/R_{\rm TAMS}$
}
\label{ta1}
$%     $$
\begin{array}{p{0.15\linewidth}cccp{0.15\linewidth}c}
\hline
            \noalign{\smallskip}
Star &  X &      Z  & \Delta \log(L)  & $\Delta \log(R)$ \\
\hline
V906 Sco A & 0.702   &0.0210 & 0.458 & 0.439    \\
           & 0.720   &0.0170 & 0.456 & 0.444    \\
           & 0.729   &0.0156 & 0.457 & 0.448    \\
           & 0.737   &0.0142 & 0.456 & 0.446    \\

            \hline
GV Car   A & 0.705   &0.0200 & 0.431 & 0.422    \\
           & 0.720   &0.0100 & 0.442 & 0.423    \\

            \noalign{\smallskip}
            \hline
\end{array}
$
\end{table*}

   \begin{figure}
%\centerline{\psfig{figure=/home/yildiz/photon/evol/BinOpClu/l_m.vs.t_r.ps,width=260bp,height=310bp}}
\centerline{\psfig{figure=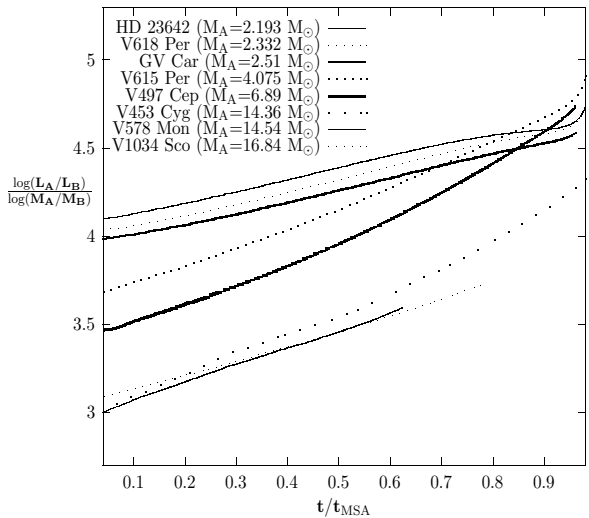,width=180bp,height=210bp}}
      \caption{The mass--luminosity relation for the DEBs in open clusters is plotted with respect to the relative age of  the primary components.
}
              {\label{f1.1}}
   \end{figure}
   \begin{figure}
%\centerline{\psfig{figure=/home/yildiz/photon/evol/BinOpClu/r_m.vs.t_r.ps,width=260bp,height=310bp}}
%\centerline{\psfig{figure=r_m.vs.t_r.ps,width=180bp,height=210bp}}
\centerline{\psfig{figure=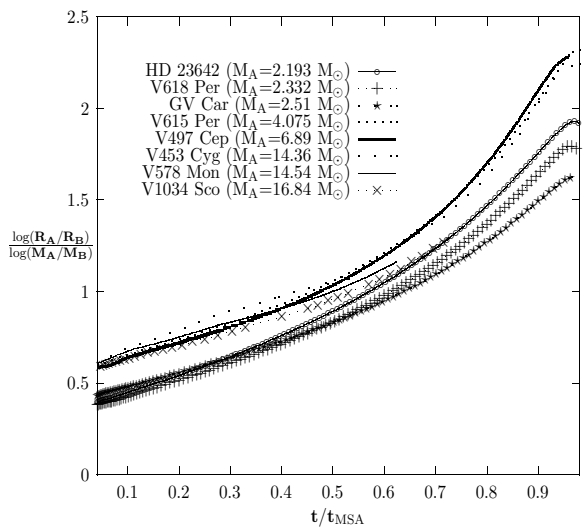,width=180bp,height=210bp}}
      \caption{The mass--radius relation for the DEBs in open clusters. 
}
              {\label{f1.1}}
   \end{figure}

{ 
The luminosity and radius of a non-rotating star are function of its mass, chemical composition 
(X and Z), and age: \\
$L=L(M,X,Z,t)$, $R=R(M,X,Z,t)$. 
%Dependence of $L$ and $R$ on any of these quantities is also function of { the rest.} 
%For the age, we can use relative age ($t/t_{\rm MSA}$).

Changes in X and Z shift the evolutionary track of a model for a given mass in HRD. Despite the variation in 
positions of ZAMS and TAMS points according to chosen values of X and Z, ratio of TAMS to ZAMS 
luminosity is independent of X, Z and stellar mass: the logarithmic luminosity difference, 
$\Delta \log(L)=L_{\rm TAMS}-L_{\rm ZAMS})/L_{\rm TAMS}$, is constant. { In Table 6, $\Delta \log(L)$ and 
$\Delta \log(R)$ for the models of V906 Sco A and GV Car A are 
given in  the fourth and the fifth columns, respectively. }
For very different combinations of X and Z, the differences are between 0.42 and 0.46: the ZAMS luminosity (radius) 
is about 44 \% less than the TAMS luminosity (radius), at least for the early-type stars.
Consider a binary system whose component masses are very different, so that the massive component is near the TAMS
and the low-mass is still near the ZAMS. For such a system, the luminosity ratio ($L_{\rm A}/L_{\rm B}$) is at a maximum.
It was at a minimum
when both of the components were near the ZAMS, and gradually increased with time. This is the case also for the 
ratio of the radii of the components. In this section, we discuss variations of radii and luminosity ratios with 
time and then develop two methods
for age determination of eclipsing binaries.

%For a binary system with different components in mass, however, luminosity-mass ratio is a measure 
%of their relative ages.  
}

%Uncertainty in age of a binary system derived from the accurate values of luminosities and radii 
%of its components depends on how these values are accurate. In most cases, as discussed ẃin the previous
%section, the uncertainty is quite high. However, it is well known that the ratio of luminosities and radii 
%of component stars in an eclipsing binary 
%are among the most accurate values derived from its light curve.

As stated above, the mass--luminosity relation is not constant and mainly depends on the stellar mass and it increases with time. 
%This is the case for stars with mass less than 10 M$_\odot$. 
%For massive stars, the mass--luminosity relation is only a function of time.
In Fig. 2, the luminosity-mass ratio \\
($l_m = \log(L_{\rm A}/L_{\rm B})/\log(M_{\rm A}/M_{\rm B}) $) obtained from the models with values given in Table 2
is plotted with respect to the relative age in MS ($t/t_{\rm MSA}$). Whereas $l_m$ increases with time, there is an inverse relation between $l_m$ and $M_{\rm A}$.
Near the ZAMS, $l_m$ varies between 3 and 4 and takes values between 4 and 5 near the TAMS. In principle, $l_m$ is very useful for predicting the age of a system
if its accuracy is good enough.

The MS life time of the primary stars $t_{\rm MSA}$, { given in the seventh column of 
Table 2}, is mainly a function of  $M_{\rm A}$, despite very different chemical composition found for the DEBs.
{ From a fit to the $M_{\rm A}$-$t_{\rm MSA}$ relation we find that}
\begin{equation}
t_{\rm MS} = \frac{3.37\times 10^9}{(M/M_\odot)^{2.122}}.
\end{equation}

\subsection{Age Determination from the Slope of the Mass--radius Relation }
The { slope of the mass--radius relation \\
$r_m = \log(R_{\rm A}/R_{\rm B})/\log(M_{\rm A}/M_{\rm B})$ is also a function of the primary mass and time.
In Fig. 3, $r_m$ is plotted with respect to $t/t_{\rm MSA}$. From the curves in Fig. 3, we determine the relative age 
of the binary systems by using the observed values of $r_m$. Multiplication of relative age of the primary component of a binary system by  $t_{\rm MSA}$ gives us age of that binary. 
The ages ($t_{\rm rm}$) of all the binaries computed with this method 
are listed in the eighth column of Table 2. 
In Fig. 4, these ages (filled circles) are plotted with respect to the ages from
fitting interior models of component stars to the observed accurate values (the fifth column in Table 2).
   \begin{figure}
%\centerline{\psfig{figure=/home/yildiz/evol/JA/r_m.vs.t_r.Tout97.ps,width=260bp,height=310bp}}
\centerline{\psfig{figure=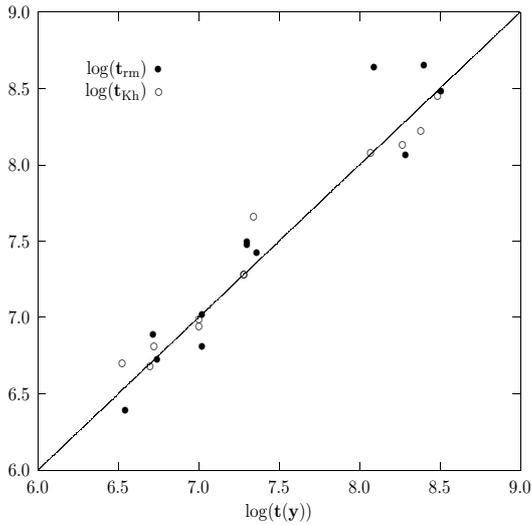,width=180bp,height=210bp}}
      \caption{The ages of binaries derived from the ratio of radii (filled circle) are plotted with respect to the ages found from fitting models. For comparison, ages given by Kharchenko {\it et al.} (2005) are also shown (circle).
}
              {\label{f1.5}}
   \end{figure}
%In Fig. 5, we compare ages found from the different methods. 
%The ages from radii ratio (filled circle) are plotted with respect to the ages from
%fitting interior models of component stars to the accurate values. 
For comparison, the ages found 
%using isochrone fitting method 
by Kharchenko {\it et al.} (2005) are also shown (circle).
There is in general a good agreement between the ages, except for V618 Per and V578 Mon. The observed value of $r_{m}$ is very high for 
V618 Per and low for V578 Mon, in comparison with the model values. For V578 Mon,  Garcia et al. (2010) report
that they find a different value for the secondary radius. The difference may significantly influence 
the observed value of $r_{m}$. 
%The largest difference occurs for 
%the V392 Car binary system, because of its very similar components.  

%The ages found from $r_{m}$--relative age relation for all the DEBs in open clusters are in very good 
%agreement with those from conventional ways are in very good agreement, except for V618 Per and V578 Mon.
%The oserved value of $r_{m}$ is very high for V618 Per and low for V578 Mon. For V578 Mon, A et al. reports
%a different value for the secondary radius, the difference may significantly influence $r_{m}$. 

The age of V1229 Tau is computed as 124 My by using $r_{m}$ from Munari et al. (2004).
%observed value of $r_{m}$ for V1229 Tau is computed as 124 My from the results given by Munari et al. (2004). 
$r_{m}$ from Southworth et al. (2005) gives age of the system as 97 My.

{ The precision of the ages found from the mass--radius curves depends on how sensitive the relative age is to 
chemical composition. 
%The relative ages found from the mass--radius curves are   
%Before going ahead further, we shall test how $r_m$-relative age relation is sensitive to chemical composition. 
In Fig. 5, $r_m$s derived from models of V618 Per A and B with different chemical compositions are 
plotted with respect to the relative age $t/t_{\rm MSA}$. For X=0.705, there is a perfect agreement between $r_m$s 
of models with Z=0.01 (thin solid line) and Z=0.02 (dotted line). We shall also compare $r_m$ values of 
models with the same Z but different X. $r_m$ of models with X=0.735 and Z=0.01 (thick solid line) 
is slightly less than 
that of the models with X=0.705 and Z=0.01. The agreement between the three curves of $r_m$ 
is good enough to suggest that relation between $r_m$ and relative age relation is almost 
independent of chemical composition.
For the precision of the ages, the sensitivity of $t_{\rm MSA}$ to chemical composition is also important.
As given in Eq. 4, $t_{\rm MSA}$ is a very strong function of stellar mass. The secondary effect of  $Z$ on $t_{\rm MSA}$ 
behaves as $t_{\rm MSA}\propto (Z/0.015)^{0.43}$ and is neglected in the present study.
}
   \begin{figure}
%\centerline{\psfig{figure=/home/yildiz/photon/evol/BinOpClu/r_m.vs.t_r.ps,width=260bp,height=310bp}}
\centerline{\psfig{figure=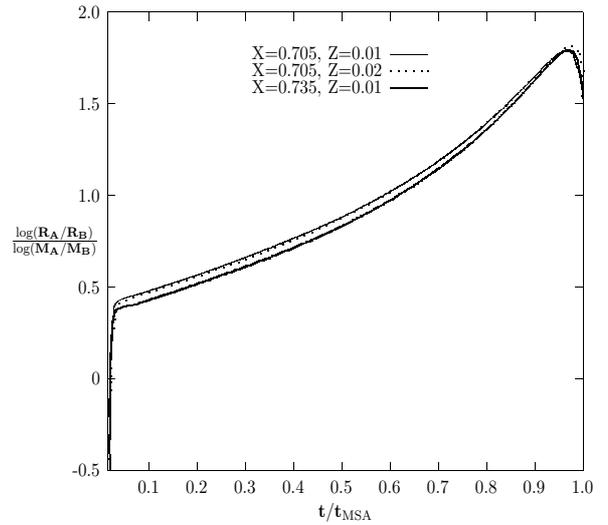,width=180bp,height=210bp}}
      \caption{The mass--radius relation for models of V618 Per with different X and Z. 
}
              {\label{f1.1}}
   \end{figure}

\subsection{Simple Expressions for the $r_m$-age Relation }
The time dependence of $r_m$ of binary systems with very different masses and chemical compositions is very similar and therefore 
can be used to estimate the age of a binary system. To do this, the time variation of $r_{m} $ is fitted with an equation of the form
\begin{equation}
r_{m} = a(t/t_{\rm MSA})^n+b.
\end{equation}
{
%The ages of the binaries are also found from the comparison of $r_{m}$ values from models given in Table 2 with 
%the observed values. We derive relative age from the curve of $r_{m}$-relative age relation for each 
%binary system. Multiplication of it by $t_{\rm MSA}$ will give us the age of that binary system. The ages for each binary system is given in the eighth column of Table 2. 

}
According to $M_{\rm A}$ of the binaries,
the curves of  $r_m$ in Fig. 3 separate essentially into two groups. The $r_m$ values of the binaries with $M_{\rm A} < $3.4 M$_\odot$ 
take place in the lower part, and those with $M_{\rm A} >$ 3.4 M$_\odot$ appear in the upper part of Fig. 3. 
Near the ZAMS of the primary component, 
$r_m$ is about 0.4 for $M_{\rm A} <$ 3.4 M$_\odot$ and 0.6 for $M_{\rm A}$ greater than 3.4 M$_\odot$. It increases with time and reaches 
a value between 1.6 and 2.4. 
%With this scheme, $r_m$ is a more sensitive function of time than $l_m$; therefore, 
It seems that $r_m$  is 
a better indicator 
of binary system age than $l_m$.
For example, age of a system with $r_m= 1$ is about half of the TAMS age of the primary component. For more accurate 
ages, two fits for the two mass ranges are made to the curves in Fig.3.  

}

{

\subsubsection{DEBs with Relatively Low-mass Primary Component ($M_{\rm A}<3.4 M_\odot$)}
%\subsubsection{DEBs with relatively low-mass primary component ($M_{\rm A}<3.4 M_\odot$)}

{ 
In order to increase the precision of Eq. (5), we derive two expressions for low and high values of  $r_{m} $.

For the three binary systems with $M_{\rm A}<3.4 M_\odot$, namely, V618 Per, V1229 Tau (HD23642) and GV Car, 
three separate fits are made to the curves in Fig. 3 by using Eq. 5. Then, the mean values of the parameters 
($a$, $b$ and $n$) given in Eq. 5 yield
%We consider model results of V618 Per,  as a representative of this group because its curve is somehow like 
%a mean of those for V1229 Tau (HD23642) and GV Car.
%Two fits are made to the curve for V618 Per given in Fig. 3 according to value of $r_{m}$. }
%
%For the binary systems whose $r_{m} $ value is about 0.4 near the ZAMS age of their primary components,
%we consider V618 Per (see Fig. 3).
\begin{equation}
%t= 0.890 {(r_{m}-0.373)^{0.758} } ~ t_{\rm MSA}.
%t= 0.941 {(r_{m}-0.39)^{0.814} } t_{\rm MSA} ~ ~ ~ {\rm for ~} r_{m} < 1.0
\frac{t}{t_{\rm MSA}}= 0.941 {(r_{m}-0.39)^{0.814} } ~ ~ ~ {\rm for ~} r_{m} < 1.0
\end{equation}
and 
\begin{equation}
%t= 0.887 {(r_{m}-0.600)^{0.4} } ~ t_{\rm MSA},
\frac{t}{t_{\rm MSA}}= 0.77 {(r_{m}-0.588)^{0.431} } ~ ~ ~{\rm for ~} r_{m} > 1.0.
\end{equation}
%The observed value of $r_{m} $ for V618 Per is 0.536 and it gives an age of 31.5 My. 
%These expressions for V618 Per is also valid for the binary systems V1229 Tau and GV Car, 
%and, in general, for the binary systems with $M_{\rm A}<3.4 M_\odot$. 
The ages of V1229 Tau and GV Car computed from these
expressions ($t_{\rm rmf}$) are given in the ninth column of Table 2. 
They are in good agreement with the ages found by other methods.
%As mentioned above, the observational value of is $r_m$ is very 

\subsubsection{DEBs with Relatively High-mass Primary Component ($M_{\rm A}>3.4 M_\odot$)}

%\subsubsection{DEBs with relatively high-mass primary component ($M_{\rm A}>3.4 M_\odot$)}
{ 
%V615 Per is considered as the typical binary system of this group.

Using Eq. 5, two fits are applied to the curves of V615 Per, V497 Cep and V453 Cyg given in Fig. 3 
according to the value of $r_{m} $. We compute the average values of the parameters $a$, $b$ and $n$.
From these values, the expressions for age as a function of $r_{m} $ are obtained: 
%For the binary systems whose $r_{m} $ value is about 0.6 near the ZAMS age of their primary components,
%we consider V615 Per and find that
\begin{equation}
%t= 0.880 {(r_{m}-0.602)^{0.730} } ~ t_{\rm MSA},
\frac{t}{t_{\rm MSA}} = 1.011 {(r_{m}-0.561)^{0.995} } ~ ~ ~ {\rm for ~} r_{m} < 1.0
\end{equation}
and
\begin{equation}
%t= 0.850 {(r_{m}-0.860)^{1/3} } ~ t_{\rm MSA},
\frac{t}{t_{\rm MSA}}= 0.613 {(r_{m}-0.852)^{0.330} } ~ ~ ~{\rm for ~} r_{m} > 1.0.
\end{equation}
}
For V615 Per,  $r_{m}= 0.747$ and Eq. (8) gives its age as 27.1 My. 
These two expressions are used to determine age of the binary systems whose $M_{\rm A}$ greater than 3.4 M$_\odot$
(see Table 2).
%The ages found from the observed values of $r_{m} $ of V615 Per and V618 Per, members of NGC 869, agree well.
%This result verifies the expressions presented above.  

{
The fitting formula given in Eqs. 6-8 are not valid for V381 Car because of 
the large mass difference between its components.
They are  also not applicable to binaries with very similar components. 
For V392 Car, for example, the mass (2.47\%) and radius (1.54\%)
differences between its components are so low that the fitting formula gives an age two times greater than 
the ages found by other methods.
The situation for DW Car is better than V392 Car. The mass (6.26\%) and radius (5.73\%)
differences between the components of DW Car are big enough to predict a reasonable value for age.
% is (6.26\% ) for mass and  5.73 \% for radius. 
For V906 Sco, however, 
there is a big difference between the radii of the components which gives so great value for $r_m$ that 
we deduce that the age is about $ t_{\rm rmf}=t_{\rm MSA}=253$My. 
 
The $r_m$ curve of V1034 Sco ($\times$ in Fig. 3), which has the highest primary mass ($M_{\rm A}$= 16.84M$_\odot$) of the stars studied, 
is slightly different from the rest. 
It is in the upper part during the early MS phase as that of the other binaries with $M_{\rm A} >$ 3.4 M$_\odot$, 
and joins later on the curves for binaries with $M_{\rm A} < $3.4 M$_\odot$. Therefore, the expressions given in 
Eqs. 6-7 shoul be used rather than Eqs. 8-9 for $M_{\rm A} >$ 15 M$_\odot$ when the relative age is greater 
than 0.5. 
%are not valid for binaries with  $M_{\rm A} >$ 15 M$_\odot$. For such binaries, somehow a mean of 
%two ages derived from Eqs. 6-7 and Eqs. 8-9 may be taken, if the relative age is greater than 0.5.  
}

The new expressions proposed above were additionally tested for selected eclipsing binaries.
The age of HD 42401
is given by { Williams (2009)} as 25 My and we find the same age ({ 25.6 My}) from the ratio of radii.
For the age of
V539 Ara Clausen {\it et al.} (1996) and Lastennet and Valls-Gabaud (2002) give 45 and 39 My, respectively. The ratio of radii 
gives the age as { 40.0} My. 

\subsection{Age Determination from the Slope of the Mass--luminosity Relation }
{
We also derive an expression to determine age from the ratio of luminosities:
\begin{equation}
t= 0.794 {(l_{m}-4.535 M_{\rm A}^{-0.154}) } 10.^{(\log M_{\rm A}-0.9)^2}~ t_{\rm MSA}.
\end{equation}
%six of them (V453 Cyg, V578 Mon, V906 Sco, V392 Car, V1034 Sco and DW Car) are listed in recent 
This expression gives acceptable ages for only the five binary systems 
(V453 Cyg, V497 Cep, V381 Car, V1034 Sco and GV Car). The reasons might be that 
%(V453 Cyg (1.12e7), V497 Cep (1.69e7), V381 Car (6.44e6), V1034 Sco (6.04e6) and GV Car(2.72e8)). 
the ratio of luminosities of component stars derived from the light curve of a binary system 
may not be as accurate as the ratio of radii. The dependence of $l_{m} $ on $M_{\rm A}$
is also as important as its time dependence (see Fig. 2).
}
\subsection{Does an $r_{m} $--relative Age Relation Implicitly Exist in the Literature?}
   \begin{figure}
%\centerline{\psfig{figure=/home/yildiz/evol/JA/r_m.vs.t_r.Tout97.ps,width=260bp,height=310bp}}
\centerline{\psfig{figure=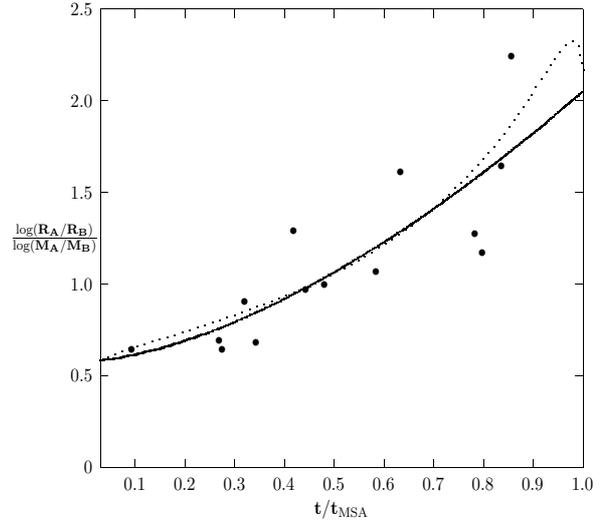,width=180bp,height=210bp}}
      \caption{The mass--radius relation for the DEBs given in Andersen (1991) with respect to the relative age of the primary components. 
The ages are taken from Pols et al. (1997). The solid line is for the curve fitted to this data 
by using Eq. 5. For comparison, $r_{m} $ of V497 Cep from models (dotted line) is also plotted.
}
              {\label{f1.1}}
   \end{figure}
The time dependence of $r_{m} $ is not considered in the previous studies in the literature. 
However, this dependence also exists in Pols et al. (1997), for example, though the authors are not aware of it.
In Fig. 6, $r_{m}$ is plotted with respect to the ages they find 
(in units of MS lifetime of the primary components)
for the well known binaries (Andersen 1991). In order to avoid complications, the binaries with $M_{\rm A}>1.5$ and 
significantly different components (difference between masses of the components is greater than 10\%) are considered.
{ The mass range is 1.56-9.245  M$_\odot$.}
The solid line shows the fitted curve for this data using Eq. 5. The dotted line represents the $r_{m} $ 
of V497 Cep, already plotted in Fig. 3. The agreement between the two curves is amazing.
}

\section{The Uncertainty in the Ages of the Binaries }
{
\begin{table*}
\caption{
The uncertainties in the ages of the binaries derived from the error boxes in the luminosity-radius diagram and
the uncertainties in the ratio of the radii of the components.
}
$%     $$
%\begin{array}{p{0.05\linewidth}lllcclllllllllll}
\begin{array}{lccccccl}
\hline 
            \noalign{\smallskip}
 Star          &     t(y) &    \Delta  t_{\rm } (y) & t_{\rm rm}(y) & \Delta  r_{\rm AB} &\Delta  t_{\rm rel} & \Delta  t_{\rm rm}(y)  & {\rm Cluster} \\
\hline
{\rm V615~Per  }    &    2 \times 10^7   & 0.61 \times 10^7    & 2.92 \times 10^7 & 0.0836 & 0.712  & 9.96 \times 10^7 & {\rm NGC~869}    \\

%{\rm V618~Per  }    &    2 \times 10^7   & 0.61 \times 10^7    & 3.15 \times 10^7 & 0.0155 & 0.040  & 2.39 \times 10^7 & {\rm NGC~869    } \\
{\rm V618~Per  }    &    2 \times 10^7   & 0.61 \times 10^7    &    --     &  --    &   --   &    --     & {\rm NGC~869    } \\

{\rm V453~Cyg  }    &   ~1.05 \times 10^7 ~& ~0.10 \times 10^7~ & ~9.76 \times 10^6 ~& 0.0193 ~& ~0.018  &~~0.02 \times 10^7~~&{\rm NGC~6871  }  \\
{\rm V1229~Tau }     &  1.93 \times 10^8  & 0.17 \times 10^8  & 1.24 \times 10^8  & 0.0045 & 0.027  & 0.19 \times 10^8 &{\rm Pleiades}\\

{\rm V578~Mon  }    &   3.50 \times 10^6  & 0.14 \times 10^6 &     --      &    --  &   --   &     --    &{\rm NGC~2244 }  \\

{\rm V906~Sco  }    &   2.52 \times 10^8  &  0.21 \times 10^8   & 2.53 \times 10^8 & 0.0241 & 0.069 & 0.17 \times 10^8 & {\rm NGC~6475  }\\

{\rm V497~Cep  }    &   2.30 \times 10^7  &  0.48 \times 10^7   & 2.17 \times 10^7 & 0.0209 & 0.053 & 0.26 \times 10^7  & {\rm NGC~7160 }\\

{\rm V381~Car  }    &   1.05 \times 10^7 &  0.11 \times 10^7  & 1.05 \times 10^7 & 0.0800 & 0.030  & 0.04 \times 10^7 & {\rm NGC~3293  }  \\

{\rm V392~Car  }    &   1.23 \times 10^8  & 0.15 \times 10^8  & 1.41 \times 10^8 & 0.0089 & 0.472  & 4.86 \times 10^8 & {\rm NGC~2516 }    \\

{\rm V1034~Sco }    &   5.51 \times 10^6  & 0.42 \times 10^6 & 5.36 \times 10^6 & 0.0080 & 0.016  & 0.16 \times 10^6 &{\rm NGC~6231 }     \\        

{\rm DW~Car    }    &   5.22 \times 10^6  &  0.30 \times 10^6 & 4.88 \times 10^6  & 0.0229 & 0.188 & 2.93 \times 10^6  &{\rm Cr~228   }      \\

{\rm GV~Car    }    &   3.22 \times 10^8  &  0.61 \times 10^8 & 3.21 \times 10^8  & 0.0406 & 0.064 & 0.27 \times 10^8  &{\rm NGC~3532}       \\
                    \noalign{\smallskip}
                    \hline
        \end{array}
        $
        \end{table*}
The ages of stars we have found involve many uncertainties, based on the model ingredients 
(e.g., opacity, nuclear reaction rates, equation of state) and the observational quantities.
It is very difficult to compute the uncertainty due to the former and therefore we take care 
in calculating the uncertainties resulting from the latter in our error analysis.
The uncertainty in the ages  ($\Delta  t$) derived from the luminosities and radii of component stars 
is the time interval in which evolutionary tracks of both components stay within the corresponding 
error boxes in luminosity-radius (L-R) diagram. The values of $\Delta  t$ for the binaries 
are given in the third column of Table 7. For a binary system whose age is derived from
its primary component, the uncertainty is also computed from that star. 
Within the context of our approach, the uncertainties are reasonably small.
 
We also compute uncertainties for the ages ($\Delta t_{\rm rm}$) derived from the ratio of the radii ($r_{\rm AB}=R_{\rm A}/R_{\rm B}$). 
{ The uncertainty in the mass ratio appearing in the denominator of $r_{\rm rm}$ is usually
very small for well-measured binary systems.}
The values of $\Delta r_{\rm AB}$ are available in the literature and given in the fifth column
of Table 7.
The error in age due to uncertainty in the ratio of the radii 
is given in the seventh column of Table 7.
The uncertainty in the relative age ($\Delta t_{\rm rel}=\Delta t/t_{\rm MSA}$) derived from the ratio of the radii ($r_{\rm AB}=R_{\rm A}/R_{\rm B}$) is computed using 
\begin{equation}
\Delta t_{\rm rel} = \frac {\Delta r_{\rm AB}}{({\partial r_{\rm AB}}{/\partial t_{\rm rel}})}.
\end{equation}
The derivative of the ratio of radii with respect to the relative age in Eq. 11  is obtained from the models 
around the relative age of each binary system. As the masses of component stars are close to each other 
the derivative is negligibly small and therefore the uncertainty in age is very large. This is the 
case for the binary systems whose components are very similar: V906 Sco, V392 Car and DW Car.
Except V615, the rest of the binary systems have very small uncertainties for their ages.
For V615 Per, the reason for the large uncertainty is the large value of $\Delta r_{\rm AB}$. 
%For V618 Per, the values of $\Delta r_{\rm AB}$ and $\Delta t_{\rm rel}$ are moderate, but 
%$\Delta  t_{\rm rm}$ is large due to long MS lifetime of the primary component.

%Except the binary systems V618 Per, V392 Car and DW Car, uncertainty in the relative ages ($\Delta t/t_{\rm MSA}$)
%is less than 0.1 for all the systems. For V392 Car and DW Car, the reason is that the masses of component 
%stars are very close to each other ($q>0.93$)
}

\section{Conclusions}
%HD 42401 (V1388 Ori) (Williams 2008,arXiv.0809.3747): $t_W$=25 My, $t_rm=25.2$ My.
%V539 Ara t=45 My (Clausen et al. 1996), t=39 My (Lastennet and Valls-Gabaud 2002)  $t_rm=39.5$ My.

Our current understanding of the evolution of our far and near universe depends on accurate age estimates for
astrophysical systems. The stars are the main constitutes of these systems,
whose ages are essentially determined from comparison of their observed and model properties.
Rotating (for DW Car A and B) and non-rotating models for the components of the eclipsing binaries, members of open clusters, are constructed to derive the ages and chemical compositions of the binaries and hence the clusters.
The largest difference between the ages we found for the eclipsing binaries and the ages given by Kharchenko {\it et al.} (2005) for the clusters
occurs for the V497 Cep binary system in NGC 7160, where the difference is about 100 \%. For the remaining systems the difference is less than 40\%.

The binary systems such as V906 Sco in NGC6475 and V453 Cyg in NGC 6871 are the most suitable systems for age determination: 
their primary components  are about to complete their MS life time and the secondary components are near or not far from the ZAMS. 
For such systems, despite equivalent solutions with very different chemical compositions, the ages are very close.
For the three different solutions with metallicity 0.014-0.017, the age of V906 Sco is in the range 252-254 My.
 
The { slope of the } mass--luminosity relation for a binary system depends on mass of its primary component $M_A$, if $M_A$ is less than about 
10 M$_\odot$. The value of \\
$l_m = \log(L_{\rm A}/L_{\rm B})/\log(M_{\rm A}/M_{\rm B}) $ near the ZAMS is 4 for $M_{\rm A}$=2.5 M$_\odot$, 3.5
for $M_{\rm A}$=6.89 M$_\odot$ and it drops to a minimum value of 3 for $M_{\rm A}$=10 M$_\odot$. For binary systems with
$M_{\rm A}$ greater than 10.M$_\odot$, $l_m=3$ near the ZAMS and reaches to 4.3 near the TAMS age of their primary component.  

The { slope of the } mass--radius relation changes with time more meaningfully than the { slope of the } mass--luminosity relation, so is more suitable for
estimating the age of the system once it is calibrated. If $r_m$=1 for a system, its age is about half of 
the MS lifetime of its primary (massive)
component. We have derived simple expressions to estimate the age of a binary system from 
the slope of mass--radius and mass--luminosity relations.
%ratio of radii and luminosities of its components.
In particular, the expressions for the slope of mass--radius relation give very reasonable results ($t_{\rm rm}$). 
%For example, $t_{\rm rm}$
%gives consistent results for the ages of V615 Per and V618 Per, 30 My, 
%despite the complicated observational values for luminosities and effective temperatures of their components.
Eqs.(6)-(9) can be used to predict age of any binary system if the radii and masses of its 
components are known. The new expressions are tested for the eclipsing binaries HD 42401 and V539 Ara and give excellent results for their ages.
The advantage of these expressions for the relative age ($t/t_{\rm MSA}$) is that they are independent of chemical composition.

\section*{Acknowledgments}

%I thank Ay\c{s}e Lahur K{\i}rtun\c{c} for { her}  suggestions which improved the language of the paper.
This work is supported by the Scientific and 
Technological Research Council of Turkey (T\"UB\.ITAK).

\def \apj#1#2{ApJ,~{#1}, #2}
\def \aj#1#2{AJ,~{#1}, #2}
\def \astroa#1{astro-ph/~{#1}}
\def \pr#1#2{Phys.~Rev.,~{#1}, #2}
\def \prt#1#2{Phys.~Rep.,~{#1}, #2}
\def \rmp#1#2{Rev. Mod. Phys.,~{#1}, #2}
\def \pt#1#2{Phys.~Today.,~{#1}, #2}
\def \pra#1#2{Phys.~Rev.,~{ A}~{#1}, #2}
\def \asap#1#2{A\&A,~{#1}, #2}
\def \aandar#1#2{A\&AR,~{#1}, #2}
\def \apss#1#2{~Ap\&SS,~{{#1}}, #2}
\def \asaps#1#2{A\&AS,~{#1}, #2}
\def \arasap#1#2{ARA\&A,~{#1}, #2}
\def \pf#1#2{Phys.~~Fluids,~{#1}, #2}
\def \apjs#1#2{ApJS,~{#1}, #2}
\def \pasj#1#2{PASJ,~{#1}, #2}
\def \mnras#1#2{MNRAS,~{#1}, #2}
\def \ibvs#1{IBVS,~No.~{#1}}

%Lastennet, E.; Valls-Gabaud, D. 2002A&A...396..551

%Clausen, J. V.; Garcia, J. M.; Gimenez, A.; Helt, B. E.; Jensen, K. S.; Suso, J.; Vaz, L. P. R. 1996A&A...308..151

% Williams, S. J. 2008, arXiv0809.3747 

\end{document}